\documentclass[aip,jcp,preprint]{revtex4-1}
\usepackage{graphicx}
\usepackage{amsmath}
\usepackage{color}
\usepackage{longtable}
\usepackage{multirow}
\DeclareGraphicsExtensions{.eps}

\begin{document}

\title{An Efficient Approach to Ab Initio Monte Carlo Simulation \footnote{The following article has been accepted by The Journal of Chemical Physics. After it is published, it will be found at http://scitation.aip.org/content/aip/journal/jcp.}
}

\author{Jeff Leiding} 
\author{Joshua D. Coe}
\email{jcoe@lanl.gov.}

\affiliation{Theoretical Division, Los Alamos National Laboratory, Los Alamos, NM, 87545}

 

\date{\today}

\begin{abstract}
We present a Nested Markov chain Monte Carlo (NMC) scheme for building equilibrium averages based on accurate potentials such as density functional theory. Metropolis sampling of a reference system, defined by an inexpensive but approximate potential, was used to substantially decorrelate configurations at which the potential of interest was evaluated, thereby dramatically reducing the number needed to build ensemble averages at a given level of precision. The efficiency of this procedure was maximized on-the-fly through variation of the reference system thermodynamic state (characterized here by its inverse temperature $\beta^0$), which was otherwise unconstrained. Local density approximation (LDA) results are presented for shocked states of argon at pressures from 4 to 60 GPa, where - depending on the quality of the reference system potential - acceptance probabilities were enhanced by factors of 1.2-28 relative to unoptimized NMC. The optimization procedure compensated strongly for reference potential shortcomings, as evidenced by significantly higher speedups when using a reference potential of lower quality. The efficiency of optimized NMC is shown to be competitive with that of standard ab initio molecular dynamics in the canonical ensemble. 
\end{abstract}

\maketitle

\section{Introduction}
Scientific computation has evolved into an immensely powerful tool, sometimes referred to as the ``third pillar" of science alongside theory and experiment.\cite{reporttopres} The popularity of atomistic simulation in particular has grown due to its fundamental character. Its predictivity can depend strongly on the intermolecular potential, however, and in this regard ab initio (AI) potentials have emerged as the de facto standard due to their combination of accuracy and generality. 

Application of AI potentials in the context of molecular dynamics (AIMD) is quite prevalent,\cite{marxhutt} partly because MD's determinism permits exploration of kinetics and mechanism in addition to thermodynamics. In certain equilibrium contexts requiring rare event sampling, however, determinism can be a liability. Such events form the basis of chemical\cite{shaw,johnson} and phase\cite{panagiotopoulos,siepmann} equilibrium under a wide range of thermodynamic conditions. Even though stochastic methods such as Markov chain Monte Carlo (MC) may prove far more efficient in such contexts, one finds (with few exceptions) ab initio Monte Carlo (AIMC) sampling to be practically non-existent. 

The challenge of combining MC with AI potentials is straightforward: schemes attempting one (or $\mathcal{O}$(1)) single-particle steps or volume adjustments between potential evaluations produce serially-correlated samples. MD configurations are correlated as well, but less so than conventional MC because MD moves are collective. As a result, uncertainty in ensemble averages can be expected to drop with the number of energy evaluations much more slowly in AIMC than in AIMD. Because the computational cost of AI energy evaluation routinely exceeds that of empirical potentials by four or more orders of magnitude, this additional expense has severely hindered the emergence of AIMC as a viable tool. It has been applied a modest number of times to small molecular clusters,\cite{cluster.jellinek,cluster.bandyopadhyay,cluster.asada,cluster.pedroza} but extension to the bulk has been rare. Wang et al.\cite{wang} made $\mathcal{O}$(N)-particle trial steps in simulation of bulk lithium, but obtained 50\% acceptance at a maximum per-particle displacement of approximately 0.05 $\AA$; AIMC can be made \emph{much} more efficient than this, as we will demonstrate below.

Nested MC (NMC) methods\cite{iftimie} lessen correlation through introduction of a reference system designed to guide MC trial steps in the spirit of ``smart" MC.\cite{rossky,rao} In NMC, a chain of \emph{M} steps is taken with a simplified and less expensive potential (henceforth, the ``reference potential"), each of which is accepted according to the Metropolis criterion based on reference system energies 
\begin{equation}
\alpha_{ij}=\min\left\{1,\frac{\pi_j^0q^0_{ji}}{\pi_i^0q^0_{ij}}\right\}.
\label{eqn:metropolis}
\end{equation}
$\alpha_{ij}$ is the probability of accepting a move from state $i$ to state $j$, $\pi_i^0$ ($\pi_j^0$) is the weight of state $i$ ($j$) in the reference distribution, and $q^0_{ij}$ is the marginal probability of proposing the $i\rightarrow j$ move. The potential of interest (henceforth, the ``full potential") is evaluated at the endpoints of the sequence, which is then accepted or rejected \emph{in toto} on the basis of a modified Metropolis criterion built from both reference and full system energies 
\begin{equation}
\alpha_{ij}=\min\left\{1,\frac{\pi_j\pi_i^0}{\pi_i\pi_j^0}\right\}.
\label{eqn:nested_acc}
\end{equation}
Quantities without superscript are analogous to those defined previously, but for the full system, and $q_{ji}/q_{ij}$ has been reexpressed in terms of $\pi_i^0/\pi_j^0$. Details and proof that this prescription recovers a Boltzmann-weighted distribution of states in the full system alone can be found elsewhere.\cite{iftimie,gelb}  For a suitably chosen reference potential, this procedure increases the distance in configuration space between full system energy evaluations \emph{without} dramatically reducing the acceptance probability of the composite step. 

Narrowing to the case of Boltzmann statistics in the canonical ensemble,
\begin{equation}
\frac{\pi_j\pi_i^0}{\pi_i\pi_j^0} = e^{-\beta\left(u_j-u_i\right) + \beta^0\left(u_j^0-u_i^0\right)},
\end{equation}
where $u$ ($u^0$) is the internal energy of the full (reference) system. For convenience we rewrite the argument of the exponential such that
\begin{align}
\Delta W &\equiv -\beta\left(u_j-u_i\right) + \beta^0\left(u_j^0-u_i^0\right) \nonumber\\
&= -\beta\left(\delta u_{ij} - \theta \delta u_{ij}^0\right), 
\end{align}
where we have introduced the scaled temperature $\theta=\beta^0/\beta$. The acceptance probability of NMC steps is now written
\begin{equation}
\label{eqn:accprob}
\alpha_{ij} = \min\left(1,e^{\Delta W\left(\theta\right)}\right).
\end{equation}

Nested AIMC has been used with some frequency\cite{naimc.kuo,McGrath1,McGrath2,McGrath1,McGrath2,gelb2,naimc.McGrath4,McGrath.mp.2006,Bandyopadhyay_tca,coe2,nakayama,McGrath3,naimc.McGrath3,Baer,bulusu,Bandyopadhyay, McGrath.pccp.2013} since the nested algorithm's inception, often with greater than $\mathcal{O}(1)$ single-particle steps per evaluation of \eqref{eqn:nested_acc}. Previously\cite{coe2} we enhanced efficiency by varying the thermodynamic state of the reference system (as defined by its pressure and temperature, for instance) so as to maximize the mean acceptance probability of nested steps, $\overline{\alpha}_{ij}$. The method was based on an exact expression for $\overline{\alpha}_{ij}$ in the limit that $i$ and $j$ are fully decorrelated, which we then evaluated by statistically sampling $\Delta W$ at a few hundred configurations drawn from the reference distribution. Because $\beta$ is a free parameter when sampling the reference distribution, we readjusted it to maximize $\overline{\alpha}_{ij}$ and then applied the reverse transformation to obtain an optimal $\beta^0$. While this produced large efficiency enhancements, the procedure was complicated and required an entirely separate simulation prior to production runs. Here we introduce a new optimization scheme that is simple, efficient, and capable of being implemented on-the-fly. 

\section{Optimization Algorithm:}
Maximization of $\overline{\alpha}_{ij}$ in the manner described above was found also to reduce the absolute value of the first two moments $\overline{\Delta W}$ and $\sigma^2(\Delta W)$ of the $\Delta W$ distribution; this is reasonable given that $\Delta W\propto\delta(0)$ (a delta distribution centered at zero) would yield unit acceptance probability. Here we reverse the emphasis by minimizing $\sigma^2(\Delta W)$ directly as a means of enhancing acceptance probability. Standard rules of variance manipulation yield
\begin{equation}
\sigma^2(\Delta W)=\beta^2(A-2B\theta+A^0\theta^2),
\label{eqn:var}
\end{equation}
where we have introduced
\begin{equation}
A\equiv\sigma^2(u_i)+\sigma^2(u_j)-2\text{ cov}(u_i,u_j)=\sigma^2(\delta u_{ij}),
\label{eqn:a}
\end{equation}
\begin{equation}
\label{eqn:a0}
A^0\equiv\sigma^2(u_i^0)+\sigma^2(u_j^0)-2\text{ cov}(u_i^0,u_j^0)=\sigma^2(\delta u^0_{ij}),
\end{equation}
and
\begin{equation}
\label{eqn:b}
B\equiv\text{cov}(u_i,u_i^0)+\text{cov}(u_j,u_j^0)-\text{cov}(u_i,u_j^0)-\text{cov}(u_i^0,u_j).
\end{equation}
Minimal variance requires that $\frac{d\sigma^2(\Delta W)}{d\theta}=0$ and $\frac{d^2\sigma^2(\Delta W)}{d\theta^2}>0$, implying that for optimal $\theta$
\begin{equation}
\theta^*\equiv\frac{(\beta^0)^*}{\beta}=\frac{B}{A^0}
\label{eqn:theta_opt}
\end{equation}
and 
\begin{equation}
A^0>0,
\label{eqn:var_min}
\end{equation}
respectively.  (`$^*$' refers to the limiting value of an optimized quantity). The temperature of interest is that of the full system, whereas that of the reference system was introduced merely for convenience; therefore variation of $\theta$ is performed at constant $\beta$. In all of our simulations we found $B>0$, in which case \eqref{eqn:var_min} amounts to the physically reasonable restriction that the reference temperature remain positive as well. Note that \eqref{eqn:theta_opt} is symmetric with respect to interchange of $i$ and $j$.

The approach introduced here can thus be summarized as follows:
\begin{enumerate} 
\item The full potential is evaluated at an initial configuration $i$, then a sequence of steps (each tested with \eqref{eqn:metropolis}) is taken in the reference system; this chain is made as long as possible up to the correlation length of the reference potential.
\item The full potential is reevaluated at $j$ and the entire sequence is accepted or rejected as a whole according to \eqref{eqn:accprob} and assuming $\theta$=1.
\item This sequence is repeated a predetermined number of times (here, 10), at which point $\theta$ is updated on the basis of \eqref{eqn:theta_opt} using the stored reference and full system energies at sampled points $i$ and $j$. 
\item The update procedure is repeated at regular intervals using aggregated statistics until $\theta$ converges to an optimal value $\theta^{opt}$ (not necessarily identical to its limiting value $\theta^*$), according to some predetermined convergence criterion.
\end{enumerate}
The sequence of full system configurations sampled during the optimization procedure does not obey detailed balance and therefore should be excluded from ensemble averages. As discussed below, this detail is of little practical consequence due to the rapidity of $\theta$ convergence. 

\section{Simulation Details}

We tested the algorithm on a system of 100 argon atoms simulated in the canonical (NVT) ensemble. States were taken from the principal Hugoniot as reported by Carpenter et al., \cite{carpenter} spanning pressures from 4 to 60 GPa and densities from 2.3 to 3.5 g/cc. A Hugoniot is defined as the locus of final states accessible upon a single shock, see Ref. \onlinecite{forbes}. The full system was characterized by finite-temperature,\cite{mermin} plane wave density functional theory (DFT) calculations in the local density approximation (LDA)\cite{lda} performed with the Vienna ab initio simulation package (VASP).\cite{vasp1,vasp2,kresse,vasp3} We employed the projector augmented wave pseudopotential method\cite{paw1,paw2} with a plane wave cutoff of 350 eV and k-point sampling at the Baldereschi mean value point.\cite{baldereschi} Convergence of thermodynamic energy and pressure were confirmed as a function of plane wave energy cutoff and Brillouin zone sampling. Electronic bands were populated according to a  Fermi-Dirac distribution with an electronic temperature equal to that of the ionic temperature. We chose LDA over a gradient-corrected functional so that we might directly compare our results with those of Carpenter et al.\cite{carpenter}

NMC results were compared to those of AIMD employing a Nos\'e-Hoover\cite{nose,hoover} thermostat with a 40 fs relaxation time. The time step was 1 fs. Two different reference systems defined by exponential-6 (EXP6) potentials were employed, one ({\bf{I}}) a ``good" match to DFT and the other ({\bf{II}}) a ``poor" one. Potential parameters are provided in the Appendix. Because NMC imposes no restriction on the reference steps other than that of detailed balance,\cite{gelb} we chose to employ hybrid Monte Carlo (HMC)\cite{duane,mehlig} simulations performed with DL\_POLY.\cite{smith} Reference system MD chains comprised 20-60 1 fs time steps (depending on the correlation time of the reference potential) taken with the time-reversible velocity Verlet integrator.\cite{swope} Initial momenta were sampled from a Maxwell-Boltzmann distribution.  

\section{Results and Discussion}

\begin{figure}
\includegraphics[width=3.5in]{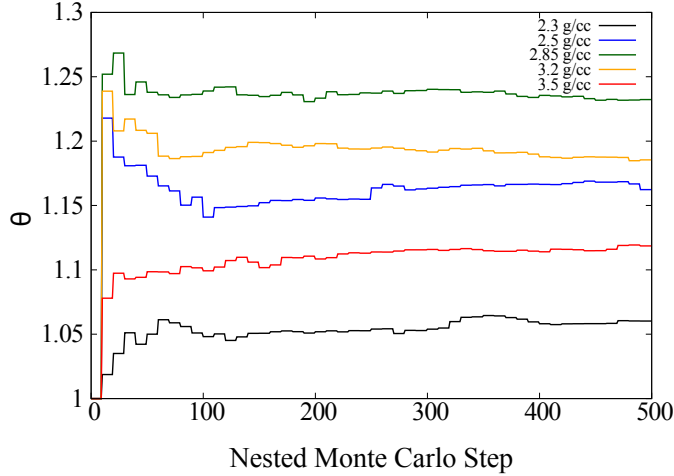}
\caption{Trace of $\theta$ for optimized NMC using reference potential {\bf{I}}. Analogous traces for {\bf{II}} are similar.}
\label{fig:thetaplot} 
\end{figure}

As shown in Figure \ref{fig:thetaplot}, scaled temperatures $\theta$ converged to their limiting values $\theta^*$ within roughly 20 updates (200 NMC steps), where $\theta^*$ was defined by five consecutive updates producing $\le$1\% change in $\theta$. We continued to update $\theta$ through 500 NMC steps (50 updates), despite there being little change after the first 5-10 updates. In order to gauge the sensitivity of performance to the degree of optimality we also tested a ``loose" convergence criterion of three consecutive updates producing $\le$5\% change in $\theta$, yielding what we will refer to as $\theta^{opt}$. The two values differed by an average of 1.9\%, with a maximum difference of 3.8\%. Results obtained using $\theta^{opt}$ will be compared to those based on $\theta^*$ below. 

\begin{figure}
\includegraphics[width=3.5in]{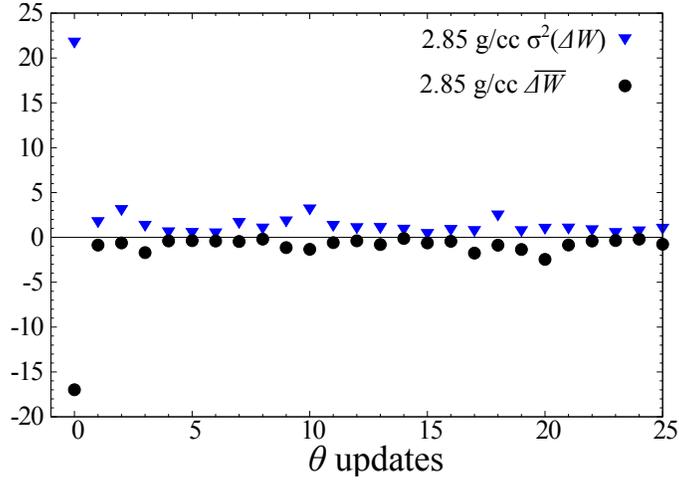}
\caption{Convergence of $\overline{\Delta W}$ and $\sigma^2(\Delta W)$ as a function of $\theta$ updates, made at 10 NMC step intervals using reference potential {\bf{II}}. The moments are essentially converged in a single adjustment.} 
\label{fig:wplot} 
\end{figure}

The optimization rate of $\Delta W$ in \eqref{eqn:accprob} was even more dramatic than that of $\theta$. Observe Figure \ref{fig:wplot}, where we show the first two moments of the $\Delta W$ distribution (based on {\bf{II}}) for $\rho=2.85$ g/cc as a function of $\theta$ adjustments. Not only did the moments change dramatically upon optimization, but they are essentially converged {\emph{after a single adjustment}}. The acceptance probability thereby increased dramatically based on a mere 11 evaluations of the full potential, after which it evolved little for the duration of the simulation. It was this rapid convergence of $\Delta W$ (a more direct indicator of acceptance probability than $\theta$) that motivated us to explore the looser convergence criterion. 

\begin{figure}[t]
\includegraphics[width=3.5in]{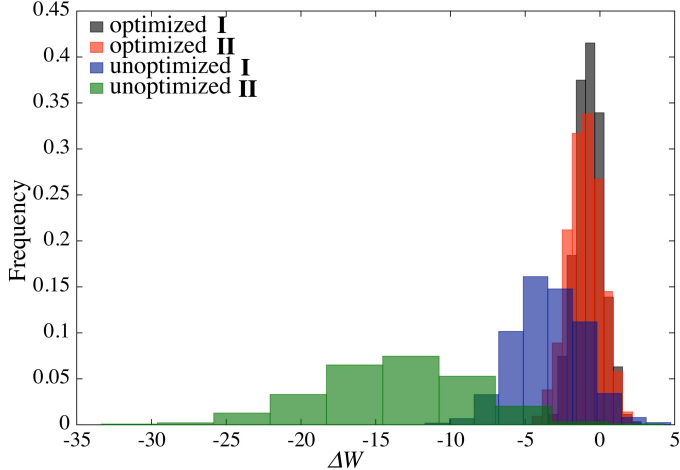}
\caption{Final $\Delta W$ distributions for $\rho$=2.85 g/cc. Optimization shifts $\overline{\Delta W}$ and $\sigma^2(\Delta W)$ to much lower values, but variation is more dramatic for reference potential {\bf{II}}.}
\label{fig:dwdist}
\end{figure}

Figure \ref{fig:dwdist} provides final histograms of $\Delta W$ for each reference potential at $\rho$=2.85 g/cc, as obtained using the fully optimized $\theta^*$. Optimization significantly shifted and narrowed the distributions built using either reference potential, but more dramatically so in the case of {\bf{II}}. Data for all other states obtained using $\theta^{opt}$ with either reference potential are provided in Table \ref{tab:dwdist}, where the same trend is observed. Table \ref{tab:dwdist} entries differ from those obtained with $\theta^*$ (not shown) by an average of only 6\%; this difference has a small impact on acceptance probability, as further discussed below. Although one would expect (rightly) the superior efficiency of {\bf{I}} on the basis of Figure \ref{fig:dwdist} and Table \ref{tab:dwdist}, the final optimized distributions are actually quite similar. This suggests that optimization compensated strongly for shortcomings in the reference potential. Such a feature is encouraging given that argon's physics are particularly simple, and for most systems one would not expect to find a reference potential at the fidelity level of {\bf{I}}. Note that potential {\bf{II}} is such a poor potential that it was unable to characterize the highest density state in this study (3.5 g/cc). The repulsive wall is too shallow, or the potential too ``soft'', and the particles are compressed beyond the potential maximum (characteristic of all EXP6 potentials) at short internuclear distances. 

\begin{table}[h]
\caption{First two moments of the $\Delta W$ distributions for optimized ($\theta^{opt}$) and unoptimized NMC for both potentials {\bf{I}} and {\bf{II}}.}
\label{tab:dwdist}
\begin{ruledtabular}
\begin{tabular}{ l  c  c  c  c c }
 $\rho$(g/$\text{cc}$) & Potential & $\overline{\Delta W}$ & $\overline{\Delta W}$  & $\sigma^2\left(\Delta W\right)$ &  $\sigma^2\left(\Delta W\right)$ \\
   &                     &  unopt   & opt     & unopt  & opt \\
 2.30   &   {\bf{I}}  &  -1.82   & -1.29 & 3.68   & 2.56  \\
 2.50  &   {\bf{I}}  &  -2.20  & -0.83   & 4.68   & 1.66\\
 2.85  &   {\bf{I}}  &  -2.62  & -0.39   & 5.23    & 0.85\\
 3.20   &   {\bf{I}}  &  -1.40  & -0.31  & 2.85    & 0.63\\
 3.50   &   {\bf{I}}  &  -0.80  & -0.32  & 1.62   & 0.63 \\ \\
 2.30   &   {\bf{II}} &  -6.89   & -1.49 & 16.27  & 3.21\\
 2.50   &   {\bf{II}} &  -10.11  & -0.99 & 25.28  & 1.89 \\
 2.85  &   {\bf{II}} &  -11.82  & -0.79  & 27.15  & 1.51\\
 3.20   &   {\bf{II}} &  -8.52   & -0.61 & 19.78   & 1.25\\
\end{tabular}
\end{ruledtabular}
\end{table}

These points are further reinforced by the data in Table \ref{tab:opt}. $\overline{\alpha}_{ij}$ and the deviation of $\theta^{opt}$ from unity largely reflect the quality of the reference potential, and both clearly illustrate the advantages of {\bf{I}} over {\bf{II}}. Optimized mean acceptances $\overline{\alpha}_{ij}^{opt}$ (obtained when using $\theta^{opt}$) were, on the other hand, much more similar than the unoptimized, reinforcing the fact that differences in reference potential quality were partially erased by the optimization procedure. Speedup factors of NMC using {\bf{II}} were considerably larger than those for {\bf{I}}. The reference temperature fell upon optimization in all cases, presumably reflecting the neglect of many-body stabilization in the reference potential. The average difference between $\overline{\alpha}_{ij}^*$ (obtained when using $\theta^*$) and $\overline{\alpha}_{ij}^{opt}$ was only 1.1\%, with a maximum of 2.8\% for {\bf{I}} at 2.3 g/cc. Relative insensitivity to full convergence of $\theta$ enables one to minimize the number of ab initio calls devoted to optimization, the results of which do not contribute to final averages (see Section II).

\begin{figure}[h]
\includegraphics[width=3.5in]{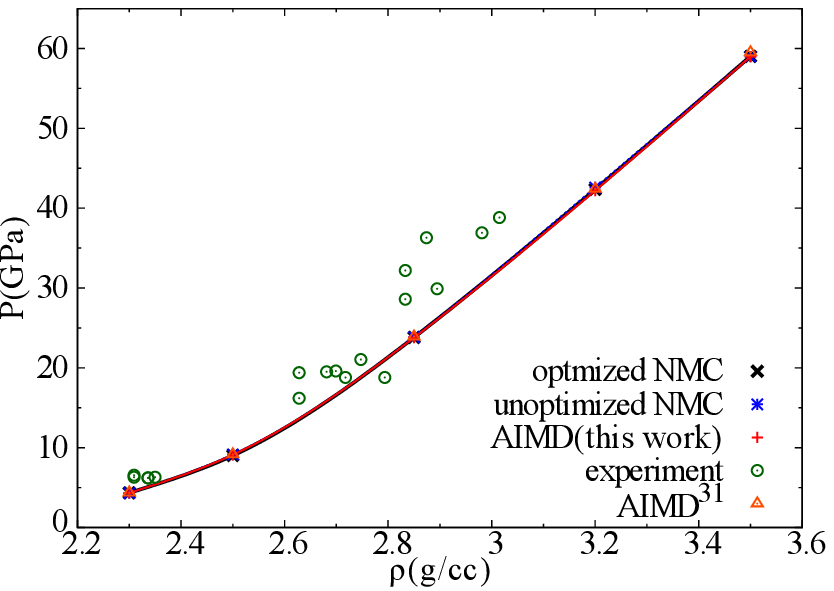}
\caption{Argon Hugoniot as computed with optimized ($\theta^{opt}$) and unoptimized NMC based on {\bf{I}}, AIMD (this work), and from experiment.\cite{nellis,vantheil} AIMD results of Carpenter et al. are included as well.\cite{carpenter} Errors on the mean were calculated as described in the text and are smaller than the symbol sizes.}
\label{fig:ArHug}
\end{figure}

\begin{table}[h]
\caption{NMC optimized and unoptimized results for both reference potentials. $\overline{\alpha}_{ij}$ is the mean acceptance probability (\%) over the course of the simulation. $^{opt}$ denotes the value obtained using the ``coarse" convergence criterion, as described in the text. Speedup is defined as $\overline{\alpha}_{ij}^{opt}/\overline{\alpha}_{ij}$.}
\label{tab:opt}
\begin{ruledtabular}
\begin{tabular}{ l  c  c  c  c  c  c }
 $\rho$(g/$\text{cc}$) & T(K) & Potential &  $\theta^{opt}$ & $\overline{\alpha}_{ij}$ & $\overline{\alpha}_{ij}^{opt}$ & $\text{Speedup}$ \\
 2.30   &  1163   &  {\bf{I}}   & 1.05 & 35.2   &42.1  & 1.20  \\
 2.50   &   2597  &  {\bf{I}}   & 1.18 & 30.2   & 53.1 & 1.76 \\
 2.85 &   7873  &  {\bf{I}}     & 1.25 & 24.9   & 67.2 &  2.70\\
 3.20   &  13690 &  {\bf{I}}   & 1.21 & 38.3  & 68.5  & 1.79 \\
 3.50   &  17700 &  {\bf{I}}   & 1.09 & 52.9  & 68.6  & 1.30\\ \\
 2.30   &  1163   &  {\bf{II}}  & 1.36 & 8.6    & 40.5   &  4.71\\
 2.50   &   2597  &  {\bf{II}}  & 1.55 & 2.8    & 48.0   &  17.14 \\
 2.85 &   7873  &   {\bf{II}}  & 1.68 & 1.9    & 53.0    &  27.89\\
 3.20   &  13690 &  {\bf{II}}  & 1.63 & 4.3   & 58.5    &  13.60\\
\end{tabular}
\end{ruledtabular}
\end{table}
The Hugoniot as calculated with unoptimized and optimized NMC and traditional AIMD is shown in Figure \ref{fig:ArHug}. Errors on the mean were estimated using standard statistical expressions that account for serial data correlation via the calculated correlation time;\cite{ant} error bars are smaller than the symbol sizes. Both versions of NMC are in excellent agreement with our AIMD results and those of Carpenter, et al.,\cite{carpenter} rendering differences among the four matters of efficiency and not of accuracy. Additional deviation from experiment is a reflection of the electronic structure level, not the sampling method.

\begin{figure}[h]
\includegraphics[width=3.5in]{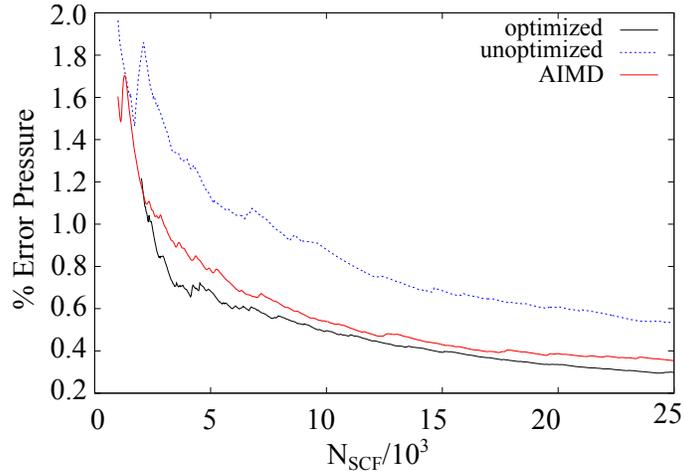}
\caption{Convergence of error in ensemble-averaged pressure for AIMD and optimized ($\theta^{opt}$) and unoptimized NMC using reference potential {\bf{I}} at $\rho$=2.85 g/cc, as a function of total SCF iterations.}
\label{fig:error}
\end{figure}

In keeping with standard practice, the converged electron density at each AIMD step was used as the initial guess for the next. The large distances separating consecutive ab initio configurations precluded such usage in NMC, meaning the latter required fewer ab initio evaluations to reach a given level of error, but more SCF iterations per energy evaluation. Therefore we compare total SCF iterations as the most reliable proxy for actual wall time, assuming the cost of reference system calculations to be relatively negligible. Convergence of the error in pressure at 2.85 g/cc for NMC based on reference potential {\bf{I}} is compared to that of AIMD in Figure \ref{fig:error}, and final results for all states are given in Table \ref{tab:stats}. Optimization consistently lowered the number of iterations required for convergence, in some cases quite sharply. These reductions would be even more dramatic for {\bf{II}}, but due to the very long simulations required for convergence of unoptimized NMC sampling using {\bf{II}}, we did not make this comparison. In many cases, optimized NMC actually required fewer SCF iterations to reach a given error than did AIMD. 

\begin{table}[h]
\caption{Statistical comparison of AIMD with optimized ($\theta^{opt}$) and unoptimized NMC based on {\bf{I}}. The number of SCF iterations reported is that required to reach an error on the mean of less than 0.5\%; errors were calculated as described in the text. SCF iterations devoted to the optimization procedure \emph{are} included in the optimized totals, because even though these points are excluded from ensemble averages, they increase the total overhead of the simulation.}
\label{tab:stats}
\begin{ruledtabular}
\begin{tabular}{ l  c  c  c  c  c  c  c  r }
 $\rho$ & \multicolumn{3}{c}{$N_{SCF}$ Energy} & \multicolumn{3}{c}{$N_{SCF}$ Pressure} \\ 
 (g/cc)  &                             AIMD         &opt                      & unopt                        & AIMD         &    opt              & unopt \\
          2.30	              &   45389    &   39554      &     48042                   &  27268       & 24559         & 26000 \\
          2.50	              &     7499    &   13362      & \textgreater 40810    &  8582         & 14807         &\textgreater 40810 \\
          2.85                       &     8048   &   4995        &  23767                      &  11245       &  9741         &\textgreater 25737 \\
          3.20	              &     4481    &   3835       &  6132                        &   7134         &  4655         &  9633\\
          3.50	              &      5044   &   4104       &   5029                       &    6292        &  4344         &   5510\\
\end{tabular}
\end{ruledtabular}
\end{table}

\section{Conclusions}
We have introduced a thermodynamically-optimal nested scheme for accelerating equilibrium AIMC sampling. Our method yields 40-70\% acceptance of $N$-particle moves at a mean per-particle displacement of 1.0 $\AA$ in dense argon; this figure is at least an order of magnitude larger than in Ref. \onlinecite{wang}, as described above. Others have attempted to optimize NMC reference potentials on-the-fly,\cite{gelb2,nakayama,bulusu} but our technique is arguably simpler with only one parameter ($\theta$) to optimize in the canonical ensemble. The method of Ref. \onlinecite{Bandyopadhyay} varied $\beta^0$ randomly between predefined values, whereas we optimized this quantity directly according to well-defined criteria. Perhaps most importantly, we found that our method compensates for shortcomings in the reference potential through correspondingly greater enhancements of acceptance probability.

Unoptimized AI NMC calculations have been performed previously in ensembles such as Gibbs\cite{McGrath2,naimc.McGrath4,McGrath.mp.2006,Baer,naimc.McGrath3,McGrath.pccp.2013} and NPT\cite{McGrath1,naimc.McGrath3}, to which our canonical optimization is readily extendable (see Appendix). We anticipate that optimization will provide even greater advantages when extended to ensembles requiring rare event sampling, such as Gibbs,\cite{panagiotopoulos} grand canonical,\cite{norman,adams} and N$_{atoms}$PT.\cite{shaw}  Such work will be presented in future publications.

\section{Acknowledgements}
We thank the Advanced Simulation and Computing (ASC) program for funding, Joel Kress for technical assistance, and Carl Greeff and Sam Shaw for helpful comments on the manuscript. LANL is operated by Los Alamos National Security, LLC, for the NNSA of the U.S. DOE under Contract No. DE-AC52-06NA25396.

\section{Appendix}
\noindent{\emph{Isothermal-isobaric ensemble}}: It is straightforward (but somewhat tedious) to show that a procedure analogous to that presented above can be applied in the isothermal-isobaric (NPT) ensemble as well. Optimal values of scaled temperature and pressure in the reference system are
\setcounter{equation}{0}
\stepcounter{equation}
\begin{equation}
\tag{A\arabic{equation}}
\theta^*=\frac{BC-DD^0}{A^0C-(D^0)^2}
\end{equation}

\stepcounter{equation}
\begin{equation}
\tag{A\arabic{equation}}
\eta^*\equiv\frac{(P^0)^*}{P}=\frac{1}{\theta^*}+\frac{A^0D-BD^0}{P(BC-DD^0)}
\end{equation}  
where $A^0$ and $B$ are as defined in the text and 

\stepcounter{equation}
\begin{equation}
\tag{A\arabic{equation}}
C\equiv\sigma^2(v_i)+\sigma^2(v_j)-2\text{ cov}(v_i,v_j)=\sigma^2(\delta v_{ij})
\end{equation}

\stepcounter{equation}
\begin{equation}
\tag{A\arabic{equation}}
D\equiv\text{cov}(u_i,v_i)+\text{cov}(u_j,v_j)-\text{ cov}(u_i,v_j)-\text{cov}(u_j,v_i)
\end{equation}

\stepcounter{equation}
\begin{equation}
\tag{A\arabic{equation}}
D^0\equiv\text{cov}(u_i^0,v_i)+\text{cov}(u_j^0,v_j)-\text{cov}(u_i^0,v_j)-\text{cov}(u_j^0,v_i).
\end{equation}
Proof and exploration of these results will be presented in a future publication.

\noindent{\em{Reference potentials}}: The EXP6 potential is of the form:
\stepcounter{equation}
\begin{equation}
\tag{A\arabic{equation}}
U^0 = A \exp\left(-r/\rho\right) - C/r^6
\end{equation}
Potentials {\bf{I}} and {\bf{II}} were taken from Ross et al.\cite{ross}
Potential {\bf{I}}: $A=58672.91$ kcal/mol, $\rho=0.3112 \AA$, $C=1578.39$ kcal/mol $\AA^6$.
Potential {\bf{II:}} $A=37517.53$ kcal/mol, $\rho=0.3225 \AA$, $C=1548.80$ kcal/mol $\AA^6$.


\begin{thebibliography}{53}%
\makeatletter
\providecommand \@ifxundefined [1]{%
 \@ifx{#1\undefined}
}%
\providecommand \@ifnum [1]{%
 \ifnum #1\expandafter \@firstoftwo
 \else \expandafter \@secondoftwo
 \fi
}%
\providecommand \@ifx [1]{%
 \ifx #1\expandafter \@firstoftwo
 \else \expandafter \@secondoftwo
 \fi
}%
\providecommand \natexlab [1]{#1}%
\providecommand \enquote  [1]{``#1''}%
\providecommand \bibnamefont  [1]{#1}%
\providecommand \bibfnamefont [1]{#1}%
\providecommand \citenamefont [1]{#1}%
\providecommand \href@noop [0]{\@secondoftwo}%
\providecommand \href [0]{\begingroup \@sanitize@url \@href}%
\providecommand \@href[1]{\@@startlink{#1}\@@href}%
\providecommand \@@href[1]{\endgroup#1\@@endlink}%
\providecommand \@sanitize@url [0]{\catcode `\\12\catcode `\$12\catcode
  `\&12\catcode `\#12\catcode `\^12\catcode `\_12\catcode `\%12\relax}%
\providecommand \@@startlink[1]{}%
\providecommand \@@endlink[0]{}%
\providecommand \url  [0]{\begingroup\@sanitize@url \@url }%
\providecommand \@url [1]{\endgroup\@href {#1}{\urlprefix }}%
\providecommand \urlprefix  [0]{URL }%
\providecommand \Eprint [0]{\href }%
\providecommand \doibase [0]{http://dx.doi.org/}%
\providecommand \selectlanguage [0]{\@gobble}%
\providecommand \bibinfo  [0]{\@secondoftwo}%
\providecommand \bibfield  [0]{\@secondoftwo}%
\providecommand \translation [1]{[#1]}%
\providecommand \BibitemOpen [0]{}%
\providecommand \bibitemStop [0]{}%
\providecommand \bibitemNoStop [0]{.\EOS\space}%
\providecommand \EOS [0]{\spacefactor3000\relax}%
\providecommand \BibitemShut  [1]{\csname bibitem#1\endcsname}%
\let\auto@bib@innerbib\@empty
\bibitem [{\citenamefont {Committee}(2005)}]{reporttopres}%
  \BibitemOpen
  \bibfield  {author} {\bibinfo {author} {\bibfnamefont {P.~I. T.~A.}\
  \bibnamefont {Committee}},\ }\href@noop {} {\enquote {\bibinfo {title}
  {Computational science: Ensuring america's competitiveness},}\ } (\bibinfo
  {year} {2005})\BibitemShut {NoStop}%
\bibitem [{\citenamefont {Marx}\ and\ \citenamefont {Hutter}(2009)}]{marxhutt}%
  \BibitemOpen
  \bibfield  {author} {\bibinfo {author} {\bibfnamefont {D.}~\bibnamefont
  {Marx}}\ and\ \bibinfo {author} {\bibfnamefont {J.}~\bibnamefont {Hutter}},\
  }\href@noop {} {\emph {\bibinfo {title} {Ab Initio Molecular Dynamics: Basic
  Theory and Advanced Methods}}}\ (\bibinfo  {publisher} {Cambridge University
  Press},\ \bibinfo {year} {2009})\BibitemShut {NoStop}%
\bibitem [{\citenamefont {Shaw}(1991)}]{shaw}%
  \BibitemOpen
  \bibfield  {author} {\bibinfo {author} {\bibfnamefont {M.~S.}\ \bibnamefont
  {Shaw}},\ }\href@noop {} {\bibfield  {journal} {\bibinfo  {journal} {J. Chem.
  Phys.}\ }\textbf {\bibinfo {volume} {94}},\ \bibinfo {pages} {7550} (\bibinfo
  {year} {1991})}\BibitemShut {NoStop}%
\bibitem [{\citenamefont {Johnson}(1999)}]{johnson}%
  \BibitemOpen
  \bibfield  {author} {\bibinfo {author} {\bibfnamefont {J.~K.}\ \bibnamefont
  {Johnson}},\ }\href@noop {} {\bibfield  {journal} {\bibinfo  {journal} {Adv.
  Chem. Phys.}\ }\textbf {\bibinfo {volume} {105}},\ \bibinfo {pages} {461}
  (\bibinfo {year} {1999})}\BibitemShut {NoStop}%
\bibitem [{\citenamefont {Panagiotopoulos}(2000)}]{panagiotopoulos}%
  \BibitemOpen
  \bibfield  {author} {\bibinfo {author} {\bibfnamefont {A.~Z.}\ \bibnamefont
  {Panagiotopoulos}},\ }\href@noop {} {\bibfield  {journal} {\bibinfo
  {journal} {J. Phys.: Condens. Matter}\ }\textbf {\bibinfo {volume} {12}},\
  \bibinfo {pages} {R25} (\bibinfo {year} {2000})}\BibitemShut {NoStop}%
\bibitem [{\citenamefont {Siepmann}(1999)}]{siepmann}%
  \BibitemOpen
  \bibfield  {author} {\bibinfo {author} {\bibfnamefont {J.~I.}\ \bibnamefont
  {Siepmann}},\ }\href@noop {} {\bibfield  {journal} {\bibinfo  {journal} {Adv.
  Chem. Phys.}\ }\textbf {\bibinfo {volume} {105}},\ \bibinfo {pages} {443}
  (\bibinfo {year} {1999})}\BibitemShut {NoStop}%
\bibitem [{\citenamefont {Jellinek}, \citenamefont {Srinivas},\ and\
  \citenamefont {Fantucci}(1998)}]{cluster.jellinek}%
  \BibitemOpen
  \bibfield  {author} {\bibinfo {author} {\bibfnamefont {J.}~\bibnamefont
  {Jellinek}}, \bibinfo {author} {\bibfnamefont {S.}~\bibnamefont {Srinivas}},
  \ and\ \bibinfo {author} {\bibfnamefont {P.}~\bibnamefont {Fantucci}},\
  }\href@noop {} {\bibfield  {journal} {\bibinfo  {journal} {Chem. Phys.
  Lett.}\ }\textbf {\bibinfo {volume} {288}},\ \bibinfo {pages} {705} (\bibinfo
  {year} {1998})}\BibitemShut {NoStop}%
\bibitem [{\citenamefont {Bandyopadhyay}, \citenamefont {Ten-No},\ and\
  \citenamefont {Iwata}(1999)}]{cluster.bandyopadhyay}%
  \BibitemOpen
  \bibfield  {author} {\bibinfo {author} {\bibfnamefont {P.}~\bibnamefont
  {Bandyopadhyay}}, \bibinfo {author} {\bibfnamefont {S.}~\bibnamefont
  {Ten-No}}, \ and\ \bibinfo {author} {\bibfnamefont {S.}~\bibnamefont
  {Iwata}},\ }\href@noop {} {\bibfield  {journal} {\bibinfo  {journal} {Mol.
  Phys.}\ }\textbf {\bibinfo {volume} {96}},\ \bibinfo {pages} {349} (\bibinfo
  {year} {1999})}\BibitemShut {NoStop}%
\bibitem [{\citenamefont {Asada}, \citenamefont {Haraguchi},\ and\
  \citenamefont {Kitaura}(2001)}]{cluster.asada}%
  \BibitemOpen
  \bibfield  {author} {\bibinfo {author} {\bibfnamefont {T.}~\bibnamefont
  {Asada}}, \bibinfo {author} {\bibfnamefont {H.}~\bibnamefont {Haraguchi}}, \
  and\ \bibinfo {author} {\bibfnamefont {K.}~\bibnamefont {Kitaura}},\
  }\href@noop {} {\bibfield  {journal} {\bibinfo  {journal} {J. Phys. Chem. A}\
  }\textbf {\bibinfo {volume} {105}},\ \bibinfo {pages} {7423} (\bibinfo {year}
  {2001})}\BibitemShut {NoStop}%
\bibitem [{\citenamefont {Pedroza}\ and\ \citenamefont
  {da~Silva}(2007)}]{cluster.pedroza}%
  \BibitemOpen
  \bibfield  {author} {\bibinfo {author} {\bibfnamefont {L.~S.}\ \bibnamefont
  {Pedroza}}\ and\ \bibinfo {author} {\bibfnamefont {A.~J.~R.}\ \bibnamefont
  {da~Silva}},\ }\href@noop {} {\bibfield  {journal} {\bibinfo  {journal}
  {Phys. Rev. B}\ }\textbf {\bibinfo {volume} {75}},\ \bibinfo {pages} {245331}
  (\bibinfo {year} {2007})}\BibitemShut {NoStop}%
\bibitem [{\citenamefont {Wang}, \citenamefont {Mitchell},\ and\ \citenamefont
  {Rikvold}(2004)}]{wang}%
  \BibitemOpen
  \bibfield  {author} {\bibinfo {author} {\bibfnamefont {S.}~\bibnamefont
  {Wang}}, \bibinfo {author} {\bibfnamefont {S.~J.}\ \bibnamefont {Mitchell}},
  \ and\ \bibinfo {author} {\bibfnamefont {P.~A.}\ \bibnamefont {Rikvold}},\
  }\href@noop {} {\bibfield  {journal} {\bibinfo  {journal} {Comput. Mater.
  Sci.}\ }\textbf {\bibinfo {volume} {29}},\ \bibinfo {pages} {145} (\bibinfo
  {year} {2004})}\BibitemShut {NoStop}%
\bibitem [{\citenamefont {Iftimie}\ \emph {et~al.}(2000)\citenamefont
  {Iftimie}, \citenamefont {Salahub}, \citenamefont {Wei},\ and\ \citenamefont
  {Schofield}}]{iftimie}%
  \BibitemOpen
  \bibfield  {author} {\bibinfo {author} {\bibfnamefont {R.}~\bibnamefont
  {Iftimie}}, \bibinfo {author} {\bibfnamefont {D.}~\bibnamefont {Salahub}},
  \bibinfo {author} {\bibfnamefont {D.}~\bibnamefont {Wei}}, \ and\ \bibinfo
  {author} {\bibfnamefont {J.}~\bibnamefont {Schofield}},\ }\href@noop {}
  {\bibfield  {journal} {\bibinfo  {journal} {J. Chem. Phys.}\ }\textbf
  {\bibinfo {volume} {113}},\ \bibinfo {pages} {4852} (\bibinfo {year}
  {2000})}\BibitemShut {NoStop}%
\bibitem [{\citenamefont {Rossky}, \citenamefont {Doll},\ and\ \citenamefont
  {Friedman}(1978)}]{rossky}%
  \BibitemOpen
  \bibfield  {author} {\bibinfo {author} {\bibfnamefont {P.~J.}\ \bibnamefont
  {Rossky}}, \bibinfo {author} {\bibfnamefont {J.~D.}\ \bibnamefont {Doll}}, \
  and\ \bibinfo {author} {\bibfnamefont {H.~L.}\ \bibnamefont {Friedman}},\
  }\href@noop {} {\bibfield  {journal} {\bibinfo  {journal} {J. Chem. Phys.}\
  }\textbf {\bibinfo {volume} {69}},\ \bibinfo {pages} {4628} (\bibinfo {year}
  {1978})}\BibitemShut {NoStop}%
\bibitem [{\citenamefont {Rao}, \citenamefont {Pangali},\ and\ \citenamefont
  {Berne}(1979)}]{rao}%
  \BibitemOpen
  \bibfield  {author} {\bibinfo {author} {\bibfnamefont {M.}~\bibnamefont
  {Rao}}, \bibinfo {author} {\bibfnamefont {C.}~\bibnamefont {Pangali}}, \ and\
  \bibinfo {author} {\bibfnamefont {B.~J.}\ \bibnamefont {Berne}},\ }\href@noop
  {} {\bibfield  {journal} {\bibinfo  {journal} {Mol. Phys.}\ }\textbf
  {\bibinfo {volume} {37}},\ \bibinfo {pages} {1773} (\bibinfo {year}
  {1979})}\BibitemShut {NoStop}%
\bibitem [{\citenamefont {Gelb}(2003)}]{gelb}%
  \BibitemOpen
  \bibfield  {author} {\bibinfo {author} {\bibfnamefont {L.~D.}\ \bibnamefont
  {Gelb}},\ }\href@noop {} {\bibfield  {journal} {\bibinfo  {journal} {J. Chem.
  Phys.}\ }\textbf {\bibinfo {volume} {118}},\ \bibinfo {pages} {7747}
  (\bibinfo {year} {2003})}\BibitemShut {NoStop}%
\bibitem [{\citenamefont {Kuo}\ \emph {et~al.}(2004)\citenamefont {Kuo},
  \citenamefont {Mundy}, \citenamefont {McGrath}, \citenamefont {Siepmann},
  \citenamefont {VandeVondele}, \citenamefont {Sprik}, \citenamefont {Hutter},
  \citenamefont {Chen}, \citenamefont {Klein}, \citenamefont {Mohamed},
  \citenamefont {Krack},\ and\ \citenamefont {Parrinello}}]{naimc.kuo}%
  \BibitemOpen
  \bibfield  {author} {\bibinfo {author} {\bibfnamefont {I.-F.~W.}\
  \bibnamefont {Kuo}}, \bibinfo {author} {\bibfnamefont {C.~J.}\ \bibnamefont
  {Mundy}}, \bibinfo {author} {\bibfnamefont {M.~J.}\ \bibnamefont {McGrath}},
  \bibinfo {author} {\bibfnamefont {J.~I.}\ \bibnamefont {Siepmann}}, \bibinfo
  {author} {\bibfnamefont {J.}~\bibnamefont {VandeVondele}}, \bibinfo {author}
  {\bibfnamefont {M.}~\bibnamefont {Sprik}}, \bibinfo {author} {\bibfnamefont
  {J.}~\bibnamefont {Hutter}}, \bibinfo {author} {\bibfnamefont
  {B.}~\bibnamefont {Chen}}, \bibinfo {author} {\bibfnamefont {M.~L.}\
  \bibnamefont {Klein}}, \bibinfo {author} {\bibfnamefont {F.}~\bibnamefont
  {Mohamed}}, \bibinfo {author} {\bibfnamefont {M.}~\bibnamefont {Krack}}, \
  and\ \bibinfo {author} {\bibfnamefont {M.}~\bibnamefont {Parrinello}},\
  }\href@noop {} {\bibfield  {journal} {\bibinfo  {journal} {J. Phys. Chem. B}\
  }\textbf {\bibinfo {volume} {108}},\ \bibinfo {pages} {12990} (\bibinfo
  {year} {2004})}\BibitemShut {NoStop}%
\bibitem [{\citenamefont {McGrath}\ \emph
  {et~al.}(2005{\natexlab{a}})\citenamefont {McGrath}, \citenamefont
  {Siepmann}, \citenamefont {Kuo}, \citenamefont {Mundy}, \citenamefont
  {VandeVondele}, \citenamefont {Hutter}, \citenamefont {Mohamed},\ and\
  \citenamefont {Krack}}]{McGrath1}%
  \BibitemOpen
  \bibfield  {author} {\bibinfo {author} {\bibfnamefont {M.~J.}\ \bibnamefont
  {McGrath}}, \bibinfo {author} {\bibfnamefont {J.~I.}\ \bibnamefont
  {Siepmann}}, \bibinfo {author} {\bibfnamefont {I.-F.~W.}\ \bibnamefont
  {Kuo}}, \bibinfo {author} {\bibfnamefont {C.~J.}\ \bibnamefont {Mundy}},
  \bibinfo {author} {\bibfnamefont {J.}~\bibnamefont {VandeVondele}}, \bibinfo
  {author} {\bibfnamefont {J.}~\bibnamefont {Hutter}}, \bibinfo {author}
  {\bibfnamefont {F.}~\bibnamefont {Mohamed}}, \ and\ \bibinfo {author}
  {\bibfnamefont {M.}~\bibnamefont {Krack}},\ }\href@noop {} {\bibfield
  {journal} {\bibinfo  {journal} {Chem. Phys. Chem.}\ }\textbf {\bibinfo
  {volume} {6}},\ \bibinfo {pages} {1894} (\bibinfo {year}
  {2005}{\natexlab{a}})}\BibitemShut {NoStop}%
\bibitem [{\citenamefont {McGrath}\ \emph
  {et~al.}(2005{\natexlab{b}})\citenamefont {McGrath}, \citenamefont
  {Siepmann}, \citenamefont {Kuo}, \citenamefont {Mundy}, \citenamefont
  {VandeVondele}, \citenamefont {Sprik}, \citenamefont {Hutter}, \citenamefont
  {Mohamed}, \citenamefont {Krack},\ and\ \citenamefont
  {Parrinello}}]{McGrath2}%
  \BibitemOpen
  \bibfield  {author} {\bibinfo {author} {\bibfnamefont {M.~J.}\ \bibnamefont
  {McGrath}}, \bibinfo {author} {\bibfnamefont {J.~I.}\ \bibnamefont
  {Siepmann}}, \bibinfo {author} {\bibfnamefont {I.-F.~W.}\ \bibnamefont
  {Kuo}}, \bibinfo {author} {\bibfnamefont {C.~J.}\ \bibnamefont {Mundy}},
  \bibinfo {author} {\bibfnamefont {J.}~\bibnamefont {VandeVondele}}, \bibinfo
  {author} {\bibfnamefont {M.}~\bibnamefont {Sprik}}, \bibinfo {author}
  {\bibfnamefont {J.}~\bibnamefont {Hutter}}, \bibinfo {author} {\bibfnamefont
  {F.}~\bibnamefont {Mohamed}}, \bibinfo {author} {\bibfnamefont
  {M.}~\bibnamefont {Krack}}, \ and\ \bibinfo {author} {\bibfnamefont
  {M.}~\bibnamefont {Parrinello}},\ }\href@noop {} {\bibfield  {journal}
  {\bibinfo  {journal} {Comp. Phys. Comm.}\ }\textbf {\bibinfo {volume}
  {169}},\ \bibinfo {pages} {289} (\bibinfo {year}
  {2005}{\natexlab{b}})}\BibitemShut {NoStop}%
\bibitem [{\citenamefont {Gelb}\ and\ \citenamefont {Carnahan}(2006)}]{gelb2}%
  \BibitemOpen
  \bibfield  {author} {\bibinfo {author} {\bibfnamefont {L.~D.}\ \bibnamefont
  {Gelb}}\ and\ \bibinfo {author} {\bibfnamefont {T.~N.}\ \bibnamefont
  {Carnahan}},\ }\href@noop {} {\bibfield  {journal} {\bibinfo  {journal}
  {Chem. Phys. Lett.}\ }\textbf {\bibinfo {volume} {417}},\ \bibinfo {pages}
  {283} (\bibinfo {year} {2006})}\BibitemShut {NoStop}%
\bibitem [{\citenamefont {McGrath}\ \emph
  {et~al.}(2006{\natexlab{a}})\citenamefont {McGrath}, \citenamefont
  {Siepmann}, \citenamefont {Kuo}, \citenamefont {Mundy}, \citenamefont
  {VandeVondele}, \citenamefont {Hutter}, \citenamefont {Mohamed},\ and\
  \citenamefont {Krack}}]{naimc.McGrath4}%
  \BibitemOpen
  \bibfield  {author} {\bibinfo {author} {\bibfnamefont {M.~J.}\ \bibnamefont
  {McGrath}}, \bibinfo {author} {\bibfnamefont {J.~I.}\ \bibnamefont
  {Siepmann}}, \bibinfo {author} {\bibfnamefont {I.-F.~W.}\ \bibnamefont
  {Kuo}}, \bibinfo {author} {\bibfnamefont {C.~J.}\ \bibnamefont {Mundy}},
  \bibinfo {author} {\bibfnamefont {J.}~\bibnamefont {VandeVondele}}, \bibinfo
  {author} {\bibfnamefont {J.}~\bibnamefont {Hutter}}, \bibinfo {author}
  {\bibfnamefont {F.}~\bibnamefont {Mohamed}}, \ and\ \bibinfo {author}
  {\bibfnamefont {M.}~\bibnamefont {Krack}},\ }\href@noop {} {\bibfield
  {journal} {\bibinfo  {journal} {J. Phys. Chem. A}\ }\textbf {\bibinfo
  {volume} {110}},\ \bibinfo {pages} {640} (\bibinfo {year}
  {2006}{\natexlab{a}})}\BibitemShut {NoStop}%
\bibitem [{\citenamefont {McGrath}\ \emph
  {et~al.}(2006{\natexlab{b}})\citenamefont {McGrath}, \citenamefont
  {Siepmann}, \citenamefont {Kuo},\ and\ \citenamefont
  {Mundy}}]{McGrath.mp.2006}%
  \BibitemOpen
  \bibfield  {author} {\bibinfo {author} {\bibfnamefont {M.~J.}\ \bibnamefont
  {McGrath}}, \bibinfo {author} {\bibfnamefont {J.~I.}\ \bibnamefont
  {Siepmann}}, \bibinfo {author} {\bibfnamefont {I.-F.~W.}\ \bibnamefont
  {Kuo}}, \ and\ \bibinfo {author} {\bibfnamefont {C.~J.}\ \bibnamefont
  {Mundy}},\ }\href@noop {} {\bibfield  {journal} {\bibinfo  {journal} {Mol.
  Phys.}\ }\textbf {\bibinfo {volume} {104}},\ \bibinfo {pages} {3619}
  (\bibinfo {year} {2006}{\natexlab{b}})}\BibitemShut {NoStop}%
\bibitem [{\citenamefont {Bandyopadhyay}(2008)}]{Bandyopadhyay_tca}%
  \BibitemOpen
  \bibfield  {author} {\bibinfo {author} {\bibfnamefont {P.}~\bibnamefont
  {Bandyopadhyay}},\ }\href@noop {} {\bibfield  {journal} {\bibinfo  {journal}
  {Theor. Chem. Acc.}\ }\textbf {\bibinfo {volume} {120}},\ \bibinfo {pages}
  {307} (\bibinfo {year} {2008})}\BibitemShut {NoStop}%
\bibitem [{\citenamefont {Coe}, \citenamefont {Sewell},\ and\ \citenamefont
  {Shaw}(2009)}]{coe2}%
  \BibitemOpen
  \bibfield  {author} {\bibinfo {author} {\bibfnamefont {J.~D.}\ \bibnamefont
  {Coe}}, \bibinfo {author} {\bibfnamefont {T.~D.}\ \bibnamefont {Sewell}}, \
  and\ \bibinfo {author} {\bibfnamefont {M.~S.}\ \bibnamefont {Shaw}},\
  }\href@noop {} {\bibfield  {journal} {\bibinfo  {journal} {J. Chem. Phys.}\
  }\textbf {\bibinfo {volume} {131}},\ \bibinfo {pages} {074105} (\bibinfo
  {year} {2009})}\BibitemShut {NoStop}%
\bibitem [{\citenamefont {Nakayama}, \citenamefont {Seki},\ and\ \citenamefont
  {Taketsugu}(2009)}]{nakayama}%
  \BibitemOpen
  \bibfield  {author} {\bibinfo {author} {\bibfnamefont {A.}~\bibnamefont
  {Nakayama}}, \bibinfo {author} {\bibfnamefont {N.}~\bibnamefont {Seki}}, \
  and\ \bibinfo {author} {\bibfnamefont {T.}~\bibnamefont {Taketsugu}},\
  }\href@noop {} {\bibfield  {journal} {\bibinfo  {journal} {J. Chem. Phys.}\
  }\textbf {\bibinfo {volume} {130}},\ \bibinfo {pages} {024107} (\bibinfo
  {year} {2009})}\BibitemShut {NoStop}%
\bibitem [{\citenamefont {McGrath}\ \emph {et~al.}(2010)\citenamefont
  {McGrath}, \citenamefont {Ghogomu}, \citenamefont {Mundy}, \citenamefont
  {Kuo},\ and\ \citenamefont {Siepmann}}]{McGrath3}%
  \BibitemOpen
  \bibfield  {author} {\bibinfo {author} {\bibfnamefont {M.~J.}\ \bibnamefont
  {McGrath}}, \bibinfo {author} {\bibfnamefont {J.~N.}\ \bibnamefont
  {Ghogomu}}, \bibinfo {author} {\bibfnamefont {C.~J.}\ \bibnamefont {Mundy}},
  \bibinfo {author} {\bibfnamefont {I.-F.~W.}\ \bibnamefont {Kuo}}, \ and\
  \bibinfo {author} {\bibfnamefont {J.~I.}\ \bibnamefont {Siepmann}},\
  }\href@noop {} {\bibfield  {journal} {\bibinfo  {journal} {Phys. Chem. Chem.
  Phys.}\ }\textbf {\bibinfo {volume} {12}},\ \bibinfo {pages} {7678} (\bibinfo
  {year} {2010})}\BibitemShut {NoStop}%
\bibitem [{\citenamefont {McGrath}\ \emph {et~al.}(2011)\citenamefont
  {McGrath}, \citenamefont {Kuo}, \citenamefont {Ghogomu}, \citenamefont
  {Mundy},\ and\ \citenamefont {Siepmann}}]{naimc.McGrath3}%
  \BibitemOpen
  \bibfield  {author} {\bibinfo {author} {\bibfnamefont {M.~J.}\ \bibnamefont
  {McGrath}}, \bibinfo {author} {\bibfnamefont {I.-F.~W.}\ \bibnamefont {Kuo}},
  \bibinfo {author} {\bibfnamefont {J.~N.}\ \bibnamefont {Ghogomu}}, \bibinfo
  {author} {\bibfnamefont {C.~J.}\ \bibnamefont {Mundy}}, \ and\ \bibinfo
  {author} {\bibfnamefont {J.~I.}\ \bibnamefont {Siepmann}},\ }\href@noop {}
  {\bibfield  {journal} {\bibinfo  {journal} {J. Phys. Chem. B}\ }\textbf
  {\bibinfo {volume} {115}},\ \bibinfo {pages} {11688} (\bibinfo {year}
  {2011})}\BibitemShut {NoStop}%
\bibitem [{\citenamefont {Baer}\ \emph {et~al.}(2011)\citenamefont {Baer},
  \citenamefont {Mundy}, \citenamefont {McGrath}, \citenamefont {Kuo},
  \citenamefont {Siepmann},\ and\ \citenamefont {Tobias}}]{Baer}%
  \BibitemOpen
  \bibfield  {author} {\bibinfo {author} {\bibfnamefont {M.~D.}\ \bibnamefont
  {Baer}}, \bibinfo {author} {\bibfnamefont {C.~J.}\ \bibnamefont {Mundy}},
  \bibinfo {author} {\bibfnamefont {M.~J.}\ \bibnamefont {McGrath}}, \bibinfo
  {author} {\bibfnamefont {I.-F.~W.}\ \bibnamefont {Kuo}}, \bibinfo {author}
  {\bibfnamefont {J.~I.}\ \bibnamefont {Siepmann}}, \ and\ \bibinfo {author}
  {\bibfnamefont {D.~J.}\ \bibnamefont {Tobias}},\ }\href@noop {} {\bibfield
  {journal} {\bibinfo  {journal} {J. Chem. Phys.}\ }\textbf {\bibinfo {volume}
  {135}},\ \bibinfo {pages} {124712} (\bibinfo {year} {2011})}\BibitemShut
  {NoStop}%
\bibitem [{\citenamefont {Bulusu}\ and\ \citenamefont
  {Fournier}(2012)}]{bulusu}%
  \BibitemOpen
  \bibfield  {author} {\bibinfo {author} {\bibfnamefont {S.}~\bibnamefont
  {Bulusu}}\ and\ \bibinfo {author} {\bibfnamefont {R.}~\bibnamefont
  {Fournier}},\ }\href@noop {} {\bibfield  {journal} {\bibinfo  {journal} {J.
  Chem. Phys.}\ }\textbf {\bibinfo {volume} {136}},\ \bibinfo {pages} {064112}
  (\bibinfo {year} {2012})}\BibitemShut {NoStop}%
\bibitem [{\citenamefont {Bandyopadhyay}(2013)}]{Bandyopadhyay}%
  \BibitemOpen
  \bibfield  {author} {\bibinfo {author} {\bibfnamefont {P.}~\bibnamefont
  {Bandyopadhyay}},\ }\href@noop {} {\bibfield  {journal} {\bibinfo  {journal}
  {Chem. Phys. Lett.}\ }\textbf {\bibinfo {volume} {556}},\ \bibinfo {pages}
  {341} (\bibinfo {year} {2013})}\BibitemShut {NoStop}%
\bibitem [{\citenamefont {McGrath}\ \emph {et~al.}(2013)\citenamefont
  {McGrath}, \citenamefont {Kuo}, \citenamefont {Ngouana~W.}, \citenamefont
  {Ghogomu}, \citenamefont {Mundy}, \citenamefont {Marenich}, \citenamefont
  {Cramer}, \citenamefont {Truhlar},\ and\ \citenamefont
  {Siepmann}}]{McGrath.pccp.2013}%
  \BibitemOpen
  \bibfield  {author} {\bibinfo {author} {\bibfnamefont {M.~J.}\ \bibnamefont
  {McGrath}}, \bibinfo {author} {\bibfnamefont {I.-F.~W.}\ \bibnamefont {Kuo}},
  \bibinfo {author} {\bibfnamefont {B.~F.}\ \bibnamefont {Ngouana~W.}},
  \bibinfo {author} {\bibfnamefont {J.~N.}\ \bibnamefont {Ghogomu}}, \bibinfo
  {author} {\bibfnamefont {C.~J.}\ \bibnamefont {Mundy}}, \bibinfo {author}
  {\bibfnamefont {A.~V.}\ \bibnamefont {Marenich}}, \bibinfo {author}
  {\bibfnamefont {C.~J.}\ \bibnamefont {Cramer}}, \bibinfo {author}
  {\bibfnamefont {D.~G.}\ \bibnamefont {Truhlar}}, \ and\ \bibinfo {author}
  {\bibfnamefont {J.~I.}\ \bibnamefont {Siepmann}},\ }\href@noop {} {\bibfield
  {journal} {\bibinfo  {journal} {Phys. Chem. Chem. Phys.}\ }\textbf {\bibinfo
  {volume} {15}},\ \bibinfo {pages} {13578} (\bibinfo {year}
  {2013})}\BibitemShut {NoStop}%
\bibitem [{\citenamefont {Carpenter}\ \emph {et~al.}(2012)\citenamefont
  {Carpenter}, \citenamefont {Root}, \citenamefont {Cochrane}, \citenamefont
  {Flicker},\ and\ \citenamefont {Mattsson}}]{carpenter}%
  \BibitemOpen
  \bibfield  {author} {\bibinfo {author} {\bibfnamefont {J.~H.}\ \bibnamefont
  {Carpenter}}, \bibinfo {author} {\bibfnamefont {S.}~\bibnamefont {Root}},
  \bibinfo {author} {\bibfnamefont {K.~R.}\ \bibnamefont {Cochrane}}, \bibinfo
  {author} {\bibfnamefont {D.~G.}\ \bibnamefont {Flicker}}, \ and\ \bibinfo
  {author} {\bibfnamefont {T.~R.}\ \bibnamefont {Mattsson}},\ }\href@noop {}
  {\enquote {\bibinfo {title} {Equation of state of argon: experiments on Z,
  density functional theory (DFT) simulations, and wide-range model},}\
  }\bibinfo {type} {Tech. Rep.}\ \bibinfo {number} {SAND2012-7991}\ (\bibinfo
  {institution} {Sandia National Laboratory},\ \bibinfo {year}
  {2012})\BibitemShut {NoStop}%
\bibitem [{\citenamefont {Forbes}(2012)}]{forbes}%
  \BibitemOpen
  \bibfield  {author} {\bibinfo {author} {\bibfnamefont {J.~W.}\ \bibnamefont
  {Forbes}},\ }\href@noop {} {\emph {\bibinfo {title} {Shock Wave Compression
  of Condensed Matter: A Primer}}}\ (\bibinfo  {publisher} {Springer},\
  \bibinfo {year} {2012})\BibitemShut {NoStop}%
\bibitem [{\citenamefont {Mermin}(1965)}]{mermin}%
  \BibitemOpen
  \bibfield  {author} {\bibinfo {author} {\bibfnamefont {N.~D.}\ \bibnamefont
  {Mermin}},\ }\href@noop {} {\bibfield  {journal} {\bibinfo  {journal} {Phys.
  Rev.}\ }\textbf {\bibinfo {volume} {137}},\ \bibinfo {pages} {A1441}
  (\bibinfo {year} {1965})}\BibitemShut {NoStop}%
\bibitem [{\citenamefont {Perdew}\ and\ \citenamefont {Zunger}(1981)}]{lda}%
  \BibitemOpen
  \bibfield  {author} {\bibinfo {author} {\bibfnamefont {J.~P.}\ \bibnamefont
  {Perdew}}\ and\ \bibinfo {author} {\bibfnamefont {A.}~\bibnamefont
  {Zunger}},\ }\href@noop {} {\bibfield  {journal} {\bibinfo  {journal} {Phys.
  Rev. B}\ }\textbf {\bibinfo {volume} {23}},\ \bibinfo {pages} {5048}
  (\bibinfo {year} {1981})}\BibitemShut {NoStop}%
\bibitem [{\citenamefont {Kresse}\ and\ \citenamefont {Hafner}(1993)}]{vasp1}%
  \BibitemOpen
  \bibfield  {author} {\bibinfo {author} {\bibfnamefont {G.}~\bibnamefont
  {Kresse}}\ and\ \bibinfo {author} {\bibfnamefont {J.}~\bibnamefont
  {Hafner}},\ }\href@noop {} {\bibfield  {journal} {\bibinfo  {journal} {Phys.
  Rev. B}\ }\textbf {\bibinfo {volume} {47}},\ \bibinfo {pages} {558} (\bibinfo
  {year} {1993})}\BibitemShut {NoStop}%
\bibitem [{\citenamefont {Kresse}\ and\ \citenamefont {Hafner}(1994)}]{vasp2}%
  \BibitemOpen
  \bibfield  {author} {\bibinfo {author} {\bibfnamefont {G.}~\bibnamefont
  {Kresse}}\ and\ \bibinfo {author} {\bibfnamefont {J.}~\bibnamefont
  {Hafner}},\ }\href@noop {} {\bibfield  {journal} {\bibinfo  {journal} {Phys.
  Rev. B}\ }\textbf {\bibinfo {volume} {49}},\ \bibinfo {pages} {14251}
  (\bibinfo {year} {1994})}\BibitemShut {NoStop}%
\bibitem [{\citenamefont {Kresse}\ and\ \citenamefont
  {Furthm\"uller}(1996)}]{kresse}%
  \BibitemOpen
  \bibfield  {author} {\bibinfo {author} {\bibfnamefont {G.}~\bibnamefont
  {Kresse}}\ and\ \bibinfo {author} {\bibfnamefont {J.}~\bibnamefont
  {Furthm\"uller}},\ }\href@noop {} {\bibfield  {journal} {\bibinfo  {journal}
  {Comput. Mater. Sci.}\ }\textbf {\bibinfo {volume} {6}},\ \bibinfo {pages}
  {15} (\bibinfo {year} {1996})}\BibitemShut {NoStop}%
\bibitem [{\citenamefont {Kresse}\ and\ \citenamefont
  {Furthm\"{u}ller}(1996)}]{vasp3}%
  \BibitemOpen
  \bibfield  {author} {\bibinfo {author} {\bibfnamefont {G.}~\bibnamefont
  {Kresse}}\ and\ \bibinfo {author} {\bibfnamefont {J.}~\bibnamefont
  {Furthm\"{u}ller}},\ }\href@noop {} {\bibfield  {journal} {\bibinfo
  {journal} {Phys. Rev. B}\ }\textbf {\bibinfo {volume} {54}},\ \bibinfo
  {pages} {11169} (\bibinfo {year} {1996})}\BibitemShut {NoStop}%
\bibitem [{\citenamefont {Bl\"{o}chl}(1994)}]{paw1}%
  \BibitemOpen
  \bibfield  {author} {\bibinfo {author} {\bibfnamefont {P.~E.}\ \bibnamefont
  {Bl\"{o}chl}},\ }\href@noop {} {\bibfield  {journal} {\bibinfo  {journal}
  {Phys. Rev. B}\ }\textbf {\bibinfo {volume} {50}},\ \bibinfo {pages} {17953}
  (\bibinfo {year} {1994})}\BibitemShut {NoStop}%
\bibitem [{\citenamefont {Kresse}\ and\ \citenamefont {Joubert}(1999)}]{paw2}%
  \BibitemOpen
  \bibfield  {author} {\bibinfo {author} {\bibfnamefont {G.}~\bibnamefont
  {Kresse}}\ and\ \bibinfo {author} {\bibfnamefont {D.}~\bibnamefont
  {Joubert}},\ }\href@noop {} {\bibfield  {journal} {\bibinfo  {journal} {Phys.
  Rev. B}\ }\textbf {\bibinfo {volume} {59}},\ \bibinfo {pages} {1758}
  (\bibinfo {year} {1999})}\BibitemShut {NoStop}%
\bibitem [{\citenamefont {Baldereschi}(1973)}]{baldereschi}%
  \BibitemOpen
  \bibfield  {author} {\bibinfo {author} {\bibfnamefont {A.}~\bibnamefont
  {Baldereschi}},\ }\href@noop {} {\bibfield  {journal} {\bibinfo  {journal}
  {Phys. Rev. B}\ }\textbf {\bibinfo {volume} {7}},\ \bibinfo {pages} {5212}
  (\bibinfo {year} {1973})}\BibitemShut {NoStop}%
\bibitem [{\citenamefont {Nos\'e}(1984)}]{nose}%
  \BibitemOpen
  \bibfield  {author} {\bibinfo {author} {\bibfnamefont {S.}~\bibnamefont
  {Nos\'e}},\ }\href@noop {} {\bibfield  {journal} {\bibinfo  {journal} {J.
  Chem. Phys.}\ }\textbf {\bibinfo {volume} {81}},\ \bibinfo {pages} {511}
  (\bibinfo {year} {1984})}\BibitemShut {NoStop}%
\bibitem [{\citenamefont {Hoover}(1985)}]{hoover}%
  \BibitemOpen
  \bibfield  {author} {\bibinfo {author} {\bibfnamefont {W.~G.}\ \bibnamefont
  {Hoover}},\ }\href@noop {} {\bibfield  {journal} {\bibinfo  {journal} {Phys.
  Rev. A}\ }\textbf {\bibinfo {volume} {31}},\ \bibinfo {pages} {1695}
  (\bibinfo {year} {1985})}\BibitemShut {NoStop}%
\bibitem [{\citenamefont {Duane}\ \emph {et~al.}(1987)\citenamefont {Duane},
  \citenamefont {Kennedy}, \citenamefont {Pendleton},\ and\ \citenamefont
  {Roweth}}]{duane}%
  \BibitemOpen
  \bibfield  {author} {\bibinfo {author} {\bibfnamefont {S.}~\bibnamefont
  {Duane}}, \bibinfo {author} {\bibfnamefont {A.~D.}\ \bibnamefont {Kennedy}},
  \bibinfo {author} {\bibfnamefont {B.~J.}\ \bibnamefont {Pendleton}}, \ and\
  \bibinfo {author} {\bibfnamefont {D.}~\bibnamefont {Roweth}},\ }\href@noop {}
  {\bibfield  {journal} {\bibinfo  {journal} {Phys. Lett. B}\ }\textbf
  {\bibinfo {volume} {195}},\ \bibinfo {pages} {216} (\bibinfo {year}
  {1987})}\BibitemShut {NoStop}%
\bibitem [{\citenamefont {Mehlig}, \citenamefont {Heermann},\ and\
  \citenamefont {Forrest}(1992)}]{mehlig}%
  \BibitemOpen
  \bibfield  {author} {\bibinfo {author} {\bibfnamefont {B.}~\bibnamefont
  {Mehlig}}, \bibinfo {author} {\bibfnamefont {D.~W.}\ \bibnamefont
  {Heermann}}, \ and\ \bibinfo {author} {\bibfnamefont {B.~M.}\ \bibnamefont
  {Forrest}},\ }\href@noop {} {\bibfield  {journal} {\bibinfo  {journal} {Phys.
  Rev. B}\ }\textbf {\bibinfo {volume} {45}},\ \bibinfo {pages} {679} (\bibinfo
  {year} {1992})}\BibitemShut {NoStop}%
\bibitem [{\citenamefont {Smith}, \citenamefont {Yong},\ and\ \citenamefont
  {Rodger}(2002)}]{smith}%
  \BibitemOpen
  \bibfield  {author} {\bibinfo {author} {\bibfnamefont {W.}~\bibnamefont
  {Smith}}, \bibinfo {author} {\bibfnamefont {C.~W.}\ \bibnamefont {Yong}}, \
  and\ \bibinfo {author} {\bibfnamefont {P.~M.}\ \bibnamefont {Rodger}},\
  }\href@noop {} {\bibfield  {journal} {\bibinfo  {journal} {Mol. Sim.}\
  }\textbf {\bibinfo {volume} {28}},\ \bibinfo {pages} {385} (\bibinfo {year}
  {2002})}\BibitemShut {NoStop}%
\bibitem [{\citenamefont {Swope}\ \emph {et~al.}(1982)\citenamefont {Swope},
  \citenamefont {Andersen}, \citenamefont {Berens},\ and\ \citenamefont
  {Wilson}}]{swope}%
  \BibitemOpen
  \bibfield  {author} {\bibinfo {author} {\bibfnamefont {W.~C.}\ \bibnamefont
  {Swope}}, \bibinfo {author} {\bibfnamefont {H.~C.}\ \bibnamefont {Andersen}},
  \bibinfo {author} {\bibfnamefont {P.~H.}\ \bibnamefont {Berens}}, \ and\
  \bibinfo {author} {\bibfnamefont {K.~R.}\ \bibnamefont {Wilson}},\
  }\href@noop {} {\bibfield  {journal} {\bibinfo  {journal} {J. Chem. Phys.}\
  }\textbf {\bibinfo {volume} {76}},\ \bibinfo {pages} {637} (\bibinfo {year}
  {1982})}\BibitemShut {NoStop}%
\bibitem [{\citenamefont {Allen}\ and\ \citenamefont {Tildesley}(1987)}]{ant}%
  \BibitemOpen
  \bibfield  {author} {\bibinfo {author} {\bibfnamefont {M.~P.}\ \bibnamefont
  {Allen}}\ and\ \bibinfo {author} {\bibfnamefont {D.~J.}\ \bibnamefont
  {Tildesley}},\ }\href@noop {} {\emph {\bibinfo {title} {Computer Simulation
  of Liquids}}}\ (\bibinfo  {publisher} {Oxford University Press},\ \bibinfo
  {year} {1987})\ \bibinfo {note} {. See section 6.4.1}\BibitemShut {NoStop}%
\bibitem [{\citenamefont {Nellis}\ and\ \citenamefont
  {Mitchell}(1980)}]{nellis}%
  \BibitemOpen
  \bibfield  {author} {\bibinfo {author} {\bibfnamefont {W.~J.}\ \bibnamefont
  {Nellis}}\ and\ \bibinfo {author} {\bibfnamefont {A.~C.}\ \bibnamefont
  {Mitchell}},\ }\href@noop {} {\bibfield  {journal} {\bibinfo  {journal} {J.
  Chem. Phys.}\ }\textbf {\bibinfo {volume} {73}},\ \bibinfo {pages} {6137}
  (\bibinfo {year} {1980})}\BibitemShut {NoStop}%
\bibitem [{\citenamefont {van Thiel}\ and\ \citenamefont
  {Alder}(1966)}]{vantheil}%
  \BibitemOpen
  \bibfield  {author} {\bibinfo {author} {\bibfnamefont {M.}~\bibnamefont {van
  Thiel}}\ and\ \bibinfo {author} {\bibfnamefont {B.~J.}\ \bibnamefont
  {Alder}},\ }\href@noop {} {\bibfield  {journal} {\bibinfo  {journal} {J.
  Chem. Phys.}\ }\textbf {\bibinfo {volume} {44}},\ \bibinfo {pages} {1056}
  (\bibinfo {year} {1966})}\BibitemShut {NoStop}%
\bibitem [{\citenamefont {Norman}\ and\ \citenamefont
  {Filinov}(1969)}]{norman}%
  \BibitemOpen
  \bibfield  {author} {\bibinfo {author} {\bibfnamefont {G.~E.}\ \bibnamefont
  {Norman}}\ and\ \bibinfo {author} {\bibfnamefont {V.~S.}\ \bibnamefont
  {Filinov}},\ }\href@noop {} {\bibfield  {journal} {\bibinfo  {journal} {High
  Temp. (USSR)}\ }\textbf {\bibinfo {volume} {7}},\ \bibinfo {pages} {216}
  (\bibinfo {year} {1969})}\BibitemShut {NoStop}%
\bibitem [{\citenamefont {Adams}(1974)}]{adams}%
  \BibitemOpen
  \bibfield  {author} {\bibinfo {author} {\bibfnamefont {D.~J.}\ \bibnamefont
  {Adams}},\ }\href@noop {} {\bibfield  {journal} {\bibinfo  {journal} {Mol.
  Phys}\ }\textbf {\bibinfo {volume} {28}},\ \bibinfo {pages} {1241} (\bibinfo
  {year} {1974})}\BibitemShut {NoStop}%
\bibitem [{\citenamefont {Ross}\ and\ \citenamefont {Alder}(1967)}]{ross}%
  \BibitemOpen
  \bibfield  {author} {\bibinfo {author} {\bibfnamefont {M.}~\bibnamefont
  {Ross}}\ and\ \bibinfo {author} {\bibfnamefont {B.}~\bibnamefont {Alder}},\
  }\href@noop {} {\bibfield  {journal} {\bibinfo  {journal} {J. Chem. Phys.}\
  }\textbf {\bibinfo {volume} {46}},\ \bibinfo {pages} {4203} (\bibinfo {year}
  {1967})}\BibitemShut {NoStop}%
\end{thebibliography}


%
\end{document}